\documentclass[aps,prl,twocolumn,amsmath,amssymb,superscriptaddress]{revtex4}
\usepackage{graphicx}
\usepackage{amssymb}
\usepackage{natbib}
\usepackage{color}

\newcommand{\beq}{\begin{eqnarray}}
\newcommand{\eeq}{\end{eqnarray}}

\begin{document}

\title{Direct observation of spin excitation anisotropy in the paramagnetic orthorhombic state of BaFe$_{2-x}$Ni$_x$As$_2$}
\author{Haoran Man}
\affiliation{Department of Physics and Astronomy,
Rice University, Houston, Texas 77005, USA}

\author{Rui Zhang}
\affiliation{Department of Physics and Astronomy, Rice University, Houston, Texas 77005, USA}

\author{J. T. Park}
\affiliation{Heinz Maier-Leibnitz Zentrum, TU M{\"u}nchen, Lichtenbergstra{\ss}e 1, D-85747 Garching, Germany
}
\author{Xingye Lu}
\affiliation{Center for Advanced Quantum Studies and Department of Physics, Beijing Normal University, Beijing 100875, China}

\author{J. Kulda}
\affiliation{Institut Laue-Langevin, 71 avenue des Martyrs,  38000 Grenoble, France}

\author{A. Ivanov}
\affiliation{Institut Laue-Langevin, 71 avenue des Martyrs,  38000 Grenoble, France}

\author{Pengcheng Dai}
\email{pdai@rice.edu} 
\affiliation{Department of Physics and Astronomy, Rice University, Houston, Texas 77005, USA}
\affiliation{Center for Advanced Quantum Studies and Department of Physics, Beijing Normal University, Beijing 100875, China}

\date{\today}

\begin{abstract}
We use transport and inelastic neutron scattering measurements to investigate single crystals of 
iron pnictide BaFe$_{2-x}$Ni$_{x}$As$_{2}$ ($x=0,0.03$), which exhibit a tetragonal-to-orthorhombic
structural transition at $T_s$ and stripe antiferromagnetic order at $T_N$ ($T_s\geq T_N$). 
 Using a tunable uniaxial pressure device, we detwin the crystals and study their transport and spin excitation properties 
at antiferromagnetic  wave vector $S_1(1,0)$ and its 90$^\circ$ rotated wave vector  $S_2(0,1)$ under different pressure conditions. We find that uniaxial pressure necessary to detwin and maintain single domain orthorhombic antiferromagnetic phase of BaFe$_{2-x}$Ni$_{x}$As$_{2}$ induces resistivity and spin excitation anisotropy at temperatures above zero pressure $T_s$. In uniaxial pressure-free detwinned sample, spin excitation anisotropy between $S_1(1,0)$ and $S_2(0,1)$ first appear in the paramagnetic orthorhombic phase below $T_s$.  These results are consistent with predictions of spin nematic theory, suggesting the absence of structural or nematic phase transition above $T_s$ in iron pnictides.
\end{abstract}

\maketitle

In the phase diagrams of high-temperature superconductors, there are many exotic ordered phases which break spatial symmetries of the underlying lattice in addition
to superconductivity \cite{keimer}. One such phase is the electronic nematic phase which breaks orientational, but not translational, symmetry 
of the underlying lattice \cite{fradkin}. For iron pnictides such as BaFe$_{2-x}$Ni$_{x}$As$_{2}$ \cite{stewart,pcdai}, there exists a structural transition 
at $T_s$, where the crystal structure changes from tetragonal to orthorhombic, followed by an antiferromagnetic (AF) transition at temperature $T_N$ 
slightly below $T_s$ ($T_N\leq T_s$) \cite{hqluo,xylu13}.  In the paramagnetic state above $T_N$, there are ample evidence for an electronic nematic 
phase from transport \cite{jhchu,matanatar,fisher,chu12,HHKuo,HHKuo15}, 
magnetic torque \cite{Kasahara}, shear-modulus \cite{anna}, scanning tunneling microscopy (STM) \cite{Rosenthal,room_T},  
angle resolved photoemission spectroscopy (ARPES) \cite{mingyi}, nuclear magnetic resonance (NMR) \cite{MFu,Dioguardi}, and neutron scattering experiments \cite{xylu14,QZhang15,YSong15,XLu16,WLZhang}. In particular, 
transport \cite{jhchu,matanatar,fisher,chu12,HHKuo,HHKuo15}, 
ARPES \cite{mingyi}, and neutron scattering \cite{xylu14,YSong15,XLu16,WLZhang} experiments on single crystals of
iron pnictides reveal that the nematic phase first appears below a characteristic temperature $T^\ast$ above $T_s$ and $T_N$, where the system is in the paramagnetic tetragonal state. The nematic phase has been suggested as a distinct phase at 
$T^\ast$ well above $T_s$ \cite{Kasahara}. 
Theoretically, it has been argued that the experimentally observed electronic nematic phase is due to spin \cite{CFang,CXu,Fernandes2011,rafael,SLiang} or orbital \cite{CCLee,Wlv} degrees
of freedom, and should only appear in the paramagnetic orthorhombic phase below $T_s$.

To understand this behavior, we note that iron pinctides exhibiting tetragonal-to-orthorhombic structural transition form twin domains below $T_s$ 
due to small mismatch of the lattice constants of the orthorhombic axes ($a$ and $b$) in the FeAs plane \cite{room_T}. To unveil the intrinsic electronic properties of 
the system, an external uniaxial pressure (stress) must be applied along the in-plane orthorhombic axis, forcing the short $b$-axis 
to align with the external pressure, and drive the twinned domain sample into a single domain suitable for electronic anisotropy measurements \cite{fisher}.
Although an externally applied uniaxial pressure can effectively change the twin-domain population, it also introduces an artificial anisotropic strain field  
that breaks the four fold rotational symmetry of the paramagnetic tetragonal phase and induces an orthorhombic lattice distortion  in iron pnictides 
above $T_s$ \cite{XLu16}. While transport, neutron diffraction, and Raman scattering measurements 
carried out under tunable uniaxial pressure on single crystals of iron pnictides suggest that resistivity anisotropy found above $T_s$ in 
transport measurements \cite{jhchu,matanatar,fisher,chu12,HHKuo,HHKuo15,anna} is likely induced by the external pressure \cite{HMan,XRen15}, much is still unclear concerning 
the nature of the nematic phase and its microscopic origin.  In particular, if the electronic nematic phase has a spin origin, one would expect that spin excitation anisotropy
at the AF ordering wave vector ${\bf Q}_{AF}=S_1(1,0)$ and 90$^\circ$ rotated wave vector $S_2(0,1)$ first appears below $T_s$ with increasing spin-spin correlations
at $S_1(1,0)$ and decreasing spin-spin correlations at $S_2(0,1)$ (Fig. 1) \cite{Fernandes2011,rafael}.
Although recent inelastic neutron scattering experiments confirm the increasing  spin-spin correlations
at $S_1(1,0)$ and decreasing spin-spin correlations at $S_2(0,1)$ in electron-doped 
iron pnictide BaFe$_{1.935}$Ni$_{0.065}$As$_2$, the measurements were carried out under an uniaxial pressure and spin excitation anisotropy first 
appears at a temperature well above $T_s$ \cite{WLZhang}.  Therefore, it is still unclear if spin excitation anisotropy above $T_s$  is induced by the applied uniaxial pressure or an intrinsic property of the spin nematic phase in iron pnictide.
 
Our BaFe$_{2-x}$Ni$_{x}$As$_{2}$ ($x=0,0.03$) single crystals were grown using self-flux method [Fig. 1(a)] \cite{ycchen}.
The crystal orientations were determined by X-ray Laue machine, and the square-shaped samples were cut for neutron scattering
and transport measurements. All samples were annealed at 800 K for 2 days to reduce defects and disorder. Transport measurements 
were carried out using a physical property measurement system (PPMS). 
We used the standard four-probe method and measured resistivity on warming with a slow rate. 
The in-plane resistivity anisotropy was measured using Montgomery method as described before \cite{HMan}. 
By taking the first derivative of the resistivity data in 
BaFe$_{1.97}$Ni$_{0.03}$As$_{2}$ [Fig. \ref{UPfig2}(f)], we can see clear split of $T_N$ and $T_s$, with $T_N \approx 109$ K and $T_s \approx 113$ K.

\begin{figure}[!htb]
\begin{center}
\includegraphics[width = 8cm]{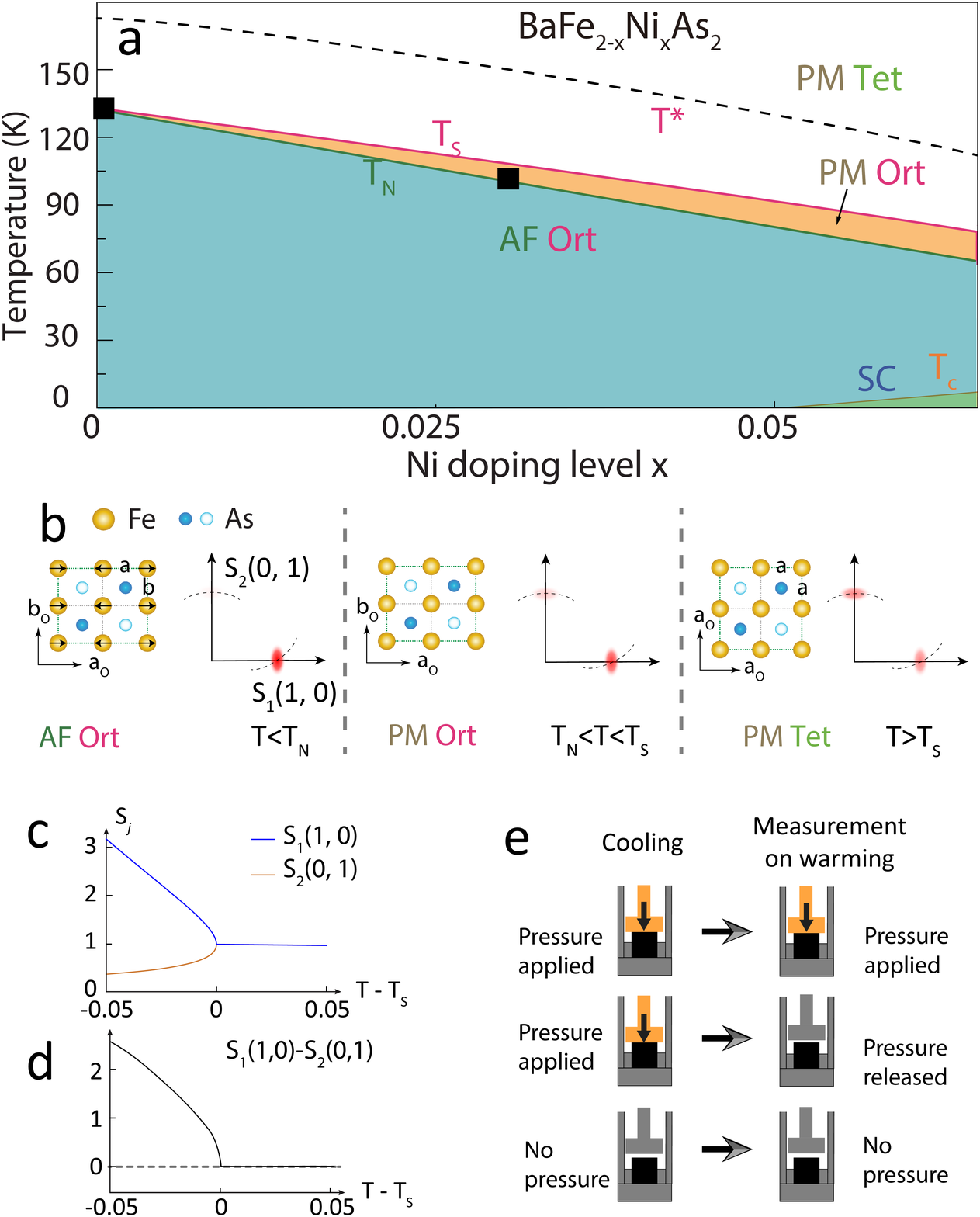}
\caption[Phase diagram, theoretical prediction, and schematics of pressure device.]{(a) 
The electronic phase diagram of BaFe$_{2-x}$Ni$ _x$As$_2$ as a function of Ni-doping as determined from previous work \cite{xylu13}. The AF orthorhombic (AF Ort), paramagnetic orthorhombic (PM Ort), paramagnetic tetragonal (PM Tet), and superconductivity (SC) phases are clearly marked. Black square points mark the Ni-doping levels measured in this paper. 
(b) Schematic illustration of Fe-As layer at different temperatures and its corresponding reciprocal space 
for temperature $T<T_N$, $T_N>T>T_s$, and $T>T_s$. The AF ordering wave vector and its 90$^\circ$ rotation are marked as 
$S_1(1,0)$ and $S_2(0,1)$, respectively. The dotted curves are 
in-plane projection of neutron scattering scan trajectories in reciprocal space. (c) 
Temperature dependence of the spin-spin correlation length at $S_1(1,0)$ (blue) and $S_2(0,1)$ (orange) across $T_s$ as 
predicted by spin nematic theory \cite{rafael}. (d) The corresponding temperature dependence of the magnetic intensity difference between 
$S_1(1,0)$ and $S_2(0,1)$. 
(e) Schematic diagrams of the in-situ device used to change pressure on the sample. 
A micrometer and a spring are used to adjust the pressure applied to the sample \cite{HMan}. The applied pressure 
can be released by a full retreat of the micrometer, leaving the sample partially detwinned at low temperature. 
}
\label{UPfig1}
\end{center}
\end{figure}

Using a specially designed tunable uniaxial pressure device \cite{HMan}, we study spin excitations at $S_1(1,0)$ and $S_2(0,1)$, and resistivity anisotropy in single domain 
 orthorhombic BaFe$_{2-x}$Ni$_{x}$As$_{2}$ [Fig. 1(a)]. The nematic order parameter 
 can be obtained by comparing the dynamic spin-spin correlation function $S({\bf Q},\omega)$ at ${\bf Q}_{AF}=S_1(1,0)$ and ${\bf Q}_2=S_2(0,1)$ in the paramagnetic 
orthorhombic ($T_s>T>T_N$) and tetragonal ($T>T_s$) phases [Figs. 1(b)-(d)] \cite{rafael}. In the stress-free state, one expects that the differences in 
 $S({\bf Q},\omega)$ at $S_1(1,0)$ and $S_2(0,1)$ would only occur below $T_s$ [Figs. \ref{UPfig1}(c) and (d)] \cite{rafael}. By measuring $S({\bf Q},\omega)$ at ${\bf Q}_{AF}$ and ${\bf Q}_2=S_2$ and comparing the outcome with transport measurements under different uniaxial pressure in BaFe$_{2-x}$Ni$_{x}$As$_{2}$,
we find that applied uniaxial pressure indeed induces spin excitation anisotropy above $T_s$, and such anisotropy only appears below $T_s$ in the stress-free sample, consistent with 
theoretical prediction \cite{rafael}.  Our transport and inelastic neutron scattering experiments thus reveal a direct correlation between spin excitation and resistivity 
anisotropy, suggesting that resistivity anisotropy and associated nematic phase has a spin origin \cite{CFang,CXu,Fernandes2011,rafael,SLiang}. 

\begin{figure}[!htb]
\begin{center}\includegraphics[width = 8cm]{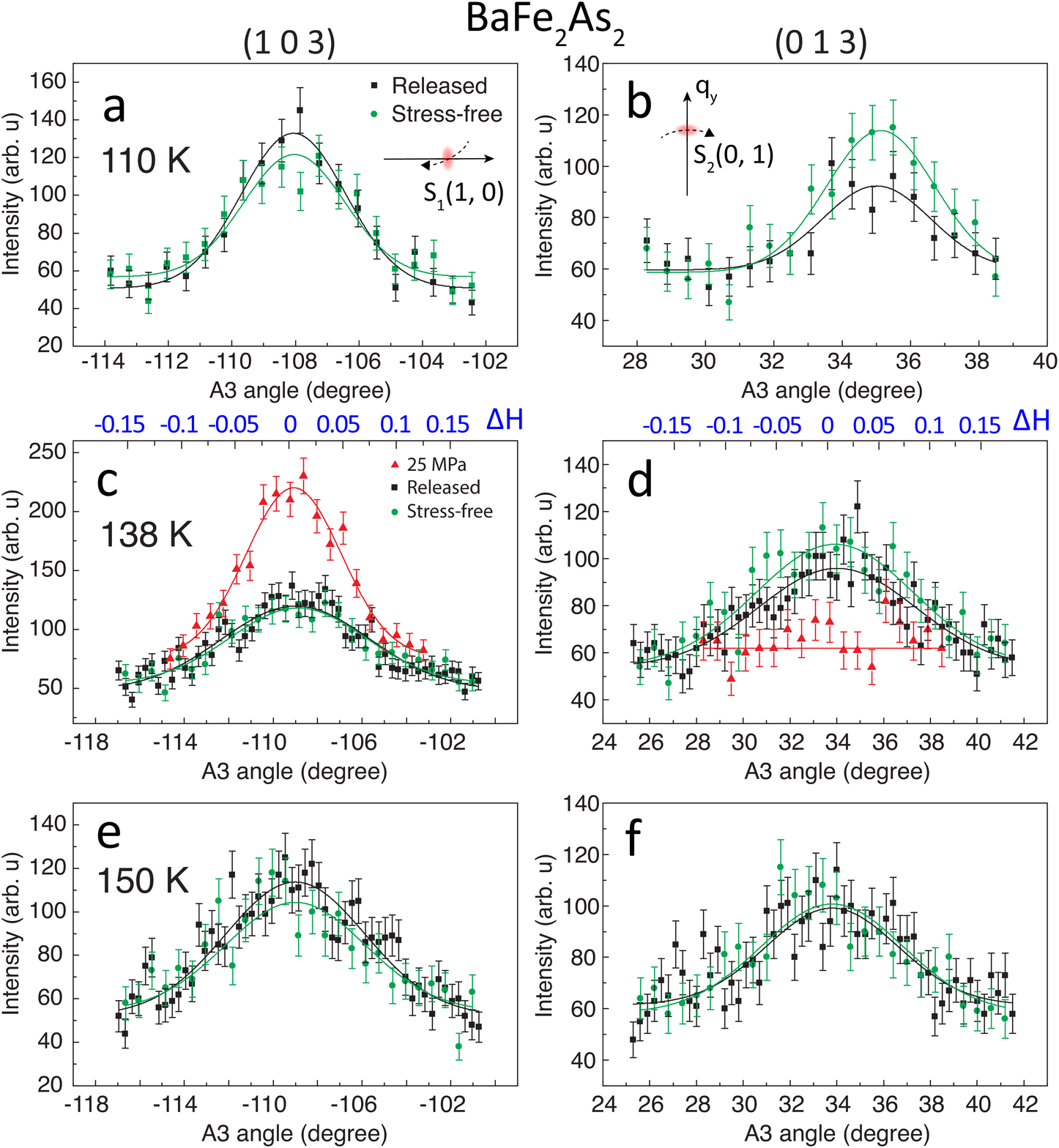}
\caption[A3 scans for BaFe$_{2}$As$_2$ at different temperature and pressure.]{
Inelastic neutron scattering measurements of 10 meV spin excitations of BaFe$_{2}$As$_2$ under different conditions at $S_1(1,0)$ and $S_2(0,1)$. 
Transverse A3 (rocking) scans at different temperature and pressure conditions. The corresponding wave vector directions in reciprocal space 
are shown in the insets of (a) and (b).
The scans are measured under 25 MPa pressure (red), pressure released (black), and stress-free (green) cases. (a,b) at 110 K ($<T_N/T_s$); (c,d) 138 K ($\approx T_N/T_s$); 
(e,f) 150 K ($>T_N/T_s$).
}
\label{UPfig3}
\end{center}
\end{figure}

We designed an in-situ mechanical device which can apply and release uniaxial pressure 
at any temperature below 300 K [Fig. \ref{UPfig1}(e)] \cite{HMan}.  
With a micrometer on top, the magnitude of the uniaxial pressure along the $b$-axis of the orthorhombic lattice is controlled by a spring compressed by the displacement of the micrometer. By applying pressure at room temperature ($\gg T_s$), cooling sample down below $T_N$, and releasing the pressure, we can measure the 
intrinsic electronic properties of iron pnictides in the AF ordered state without external pressure (stress-free).
Three types of measurements are carried out: 
\begin{enumerate}
\item Pressure applied: unaixial pressure sufficient to detwin the sample is applied during the entire measurement. Both intrinsic and pressure induced 
effect will contribute to measured transport and spin excitation anisotropy. 
\item Pressure released: a uniaxial pressure is applied on cooling from room temperature to base temperature ($\ll T_N$). The pressure is then released 
at base temperature. Transport and spin excitation measurements were carried out on warming, where the sample remains partially detwinned and only 
intrinsic electronic difference in the orthorhombic state contribute to measured transport and spin excitation anisotropy.
\item No Pressure: No uniaxial pressure is applied to the sample and the sample remains in the twinned state below $T_s$ and $T_N$. 
If twin domains are equally distributed, there should be no transport and spin excitation anisotropy.
\end{enumerate}

Our inelastic neutron scattering experiments were carried out at the IN-8 triple-axis spectrometer using a multi-analyzer detector system, Institut Laue-Langevin (ILL), Grenoble, France \cite{kulda}. For inelastic neutron scattering experiments, 
 annealed square-shaped single crystals of BaFe$_2$As$_2$ ($\sim$220 mg) or BaFe$_{1.97}$Ni$_{0.03}$As$_2$ ($\sim$200 mg) were mounted 
on the sample stick specially designed for applying uniaxial pressure (along $b$-axis) inside an orange cryostat \cite{HMan}. The momentum transfer ${\bf Q}$ in three-dimensional reciprocal space in \AA$^{-1}$ is defined as $\textbf{Q}=H\textbf{a}^\ast+K\textbf{b}^\ast+L\textbf{c}^\ast$, where $H$, $K$, and $L$ are Miller indices and 
${\bf a}^\ast=\hat{{\bf a}}2\pi/a$, ${\bf b}^\ast=\hat{{\bf b}}2\pi/b$, ${\bf c}^\ast=\hat{{\bf c}}2\pi/c$ with $a\approx b\approx 5.549$ \AA\ and 
$c\approx 12.622$ \AA\ \cite{xylu13}.
In the AF ordered state of a fully detwinned sample, the AF Bragg peaks occurs at $(\pm 1,0,L)$ ($L=1,3,5,\cdots$) positions in reciprocal space and are absent at $(0,\pm 1,L)$. The sample was aligned in the $[H,0,L]$ scattering plane. With goniometer below the orange cryostat and with extra coverage provided by the flat-cone setup on IN8 \cite{kulda}, we can access both $(1,0,3)$ and $(0,1,3)$ around 10 meV and $(1,0,5)$, $(0,1,5)$ magnetic Bragg peak positions at 0 meV.

\begin{figure}[!htb]
\begin{center}
\includegraphics[width = 6cm]{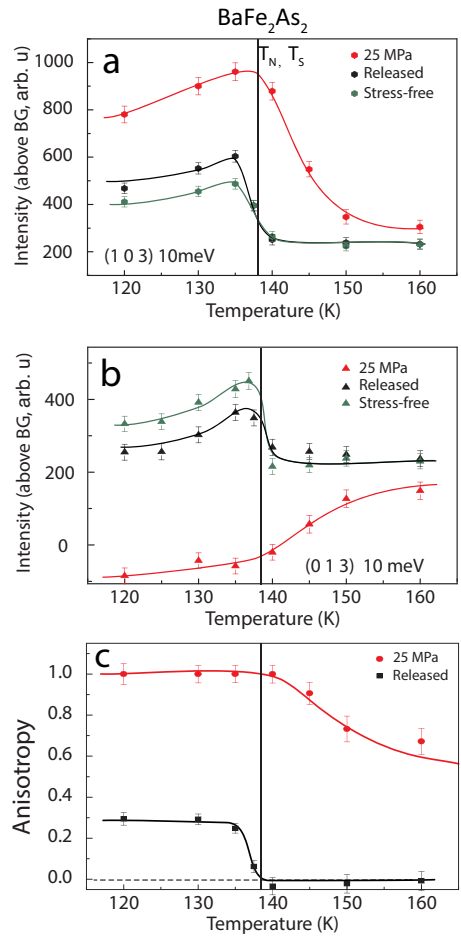}
\caption[Neutron scattering experiment for BaFe$_{2}$As$_2$.]{Temperature dependence of spin excitations at 10 meV in BaFe$_{2}$As$_2$
under 25 MPa uniaxial pressure, pressure released, and stress-free conditions at (a) AF wave vector $S_1(1,0)=(1,0,3)$; 
(b) $S_1(0,1)=(0,1,3)$. The negative scattering is due to imperfect background scattering subtraction.
(c) Temperature dependence of spin excitation anisotropy under 25 MPa pressure and pressure-released. 
}
\label{UPfig4}
\end{center}
\end{figure}

We have collected neutron scattering data under three different conditions: (1) under 22-25 MPa uniaxial pressure (pressured); 
(2) pressure released at 10 K and no pressure measurements on warming (released); and (3) no pressure at all temperatures (stress-free) [
Figure \ref{UPfig1}(e)]. For each scenario, spin excitations at wave vectors 
$(1,0,3)$ and $(0,1,3)$ are measured in the same warm-up cycle. We first test if spin excitation anisotropy seen above $T_N/T_s$ in
uniaxial pressured BaFe$_2$As$_2$ \cite{xylu14} also exists in pressure released situation.
Figures \ref{UPfig3}(a) and \ref{UPfig3}(b) compare transverse scans of spin excitations at 10 meV and 110 K ($<T_N$) for pressure released 
and stress-free cases at 
wave vectors $S_1(1,0)$ and $S_2(0,1)$, respectively.  Assuming spin excitations are isotropic in stress-free situation, we find 
clear spin excitation anisotropy in pressure released situation, consistent with previous elastic scattering measurements below $T_N$ \cite{HMan}.
On warming to 138 at $T_N/T_s$, spin excitations in the pressured case double that of the stress-free case and
only exist at $S_1(1,0)$, consistent with a fully pressure-induced detwinned state [Figs. \ref{UPfig3}(c) and \ref{UPfig3}(d)] \cite{xylu14}. 
For comparison, transverse scans in pressure released and stress-free cases are indistinguishable.
Upon further warming up to 150 K above $T_N/T_s$, we again find no spin excitation anisotropy at $S_1(1,0)$ and $S_2(0,1)$ [Figs. \ref{UPfig3}(e) and \ref{UPfig3}(f)],
 suggesting that spin excitation anisotropy only appears below $T_N/T_s$ in pressure released case.

\begin{figure}[!htb]
\begin{center}
\includegraphics[width = 8cm]{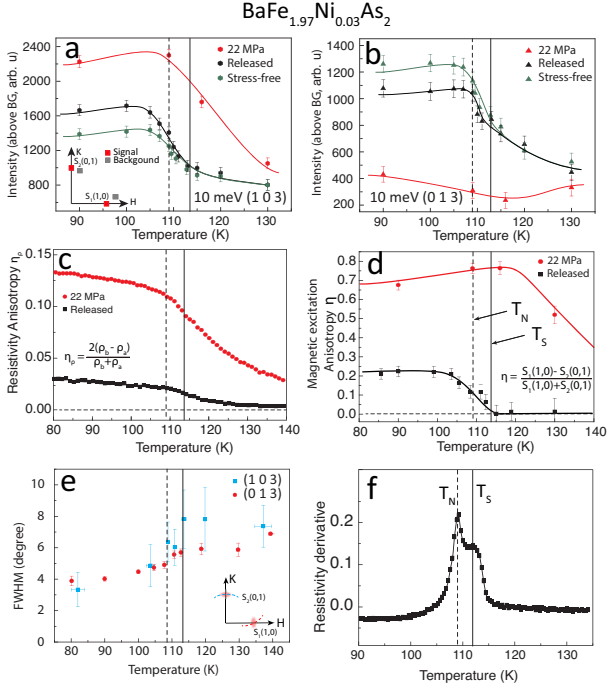}
\caption[Summary of transport and neutron scattering of BaFe$_{1.97}$Ni$_{0.03}$As$_2$.]{Temperature dependence of resistivity and spin excitations anisotropy for BaFe$_{1.97}$Ni$_{0.03}$As$_2$. (a) Temperature dependence of the 10 meV spin excitations under 22 MPa, released, and stress-free conditions at $S_1(1,0)$.  (b) Similar measurements at $S_2(0,1)$.
Temperature dependence of the (c) resistivity and, (d) spin excitation anisotropy under 22 MPa and pressure released conditions.
(e) Temperature dependence of the FWHM of the 10 meV spin excitations at $S_1(1,0)$ and $S_2(0,1)$. The scan directions are marked by dotted curve in the inset. 
(f) Temperature dependence of the first derivative of resistivity, where $T_N$ and $T_s$ are marked as vertical lines.   
}
\label{UPfig2}
\end{center}
\end{figure}

To confirm these results, we carried out spin excitation measurements in these three pressure conditions using long counting time at the peak center and subtracted background points. 
 Figures \ref{UPfig4}(a) and \ref{UPfig4}(b) show the background subtracted peak intensities at  $S_1(1,0)=(1,0,3)$ and $S_2(0,1)=(0,1,3)$, respectively. 
In case one under 25 MPa uniaxial pressure (red), the large spin excitation anisotropy at 
$S_1(1,0)$ and $S_2(0,1)$ below $T_N/T_s$ persists to temperatures well above $T_N/T_s$, consistent with earlier results \cite{xylu14}.
For case two pressure released measurements (black), while the spin excitation anisotropy between 
$S_1(1,0)=(1,0,3)$ and $S_2(0,1)=(0,1,3)$ becomes much smaller compared with stress-free case, it is still present 
below 138 K but vanishes above 138 K at both wave vectors. By normalizing pressured and released data with stress-free measurements at $S_1(1,0)$ and $S_2(0,1)$, 
we can estimate the spin excitations anisotropy $\eta$ using $\eta = (I_{(1,0)}-I_{(0,1)})/(I_{(1,0)}+I_{(0,1)})$
, where $I_{(1,0)}$ and $I_{(0,1)}$ are spin excitations at $S_1(1,0)$ and $S_2(0,1)$, respectively. 
For a fully detwinned sample in the stress-free AF ordered state, only $I_{(1,0)}$ should be present and $\eta = 1$.
In the stress-free paramagnetic tetragonal state, we expect $I_{(1,0)} = I_{(0,1)}$ and $\eta = 0$. 
Figure \ref{UPfig4}(c) indicates that uniaxial pressure is necessary to maintain 100\% detwinned state in BaFe$_2$As$_2$.  While spin excitations still have
anisotropy in the pressured released case below $T_N/T_s$, the anisotropy completely vanishes above $T_N/T_s$.  These measurements confirm that spin excitation anisotropy above $T_N/T_s$ are entirely induced by externally applied uniaxial pressure.

Having established the vanishing spin excitation anisotropy in the paramagnetic state of 
BaFe$_2$As$_2$, where 
$T_N\approx T_s$ without applied uniaxial pressure, 
 it is interesting to ask if spin excitations can be anisotropic in the stress-free paramagnetic orthorhombic nematic phase as predicted by spin nematic theory \cite{CFang,CXu,Fernandes2011,rafael,SLiang}. For this experiment, we chose BaFe$_{1.97}$Ni$_{0.03}$As$_2$ because of its separated $T_N$ 
and $T_s$ ($>T_N$) [Fig. \ref{UPfig2}(f)] \cite{xylu13}. 
Figures \ref{UPfig2}(a) and \ref{UPfig2}(b) show temperature dependence of the magnetic scattering at 10 meV for 
$S_1(1,0)=(1,0,3)$ and $S_2(0,1)=(0,1,3)$, respectively. Similar to measurement on BaFe$_2$As$_2$, spin excitation anisotropy under 22 MPa uniaxial pressure (red) extends
to temperatures well above $T_s$ [Figs. \ref{UPfig2}(a),(b),(d)]. However, the pressure released data (black) is much closer to 
stress-free data on approaching $T_s$. For temperature above $T_s$, there are no detectable differences between $S_1(1,0)$ and $S_2(0,1)$, as seen in Fig. \ref{UPfig2}(d).
Therefore, spin excitations exhibit a weak anisotropy in the paramagnetic orthorhombic phase of BaFe$_{1.97}$Ni$_{0.03}$As$_2$, consistent with theoretical expectation for
a spin excitation driven nematic phase \cite{CFang,CXu,Fernandes2011,rafael,SLiang}. For comparison, Figure \ref{UPfig2}(c) shows temperature 
dependence of the resistivity anisotropy obtained under  22 MPa uniaxial pressure (red) and pressure released (black) conditions. 
For pressure released case, we find no evidence of time dependent relaxation of the resistivity anisotropy within several hours.
  Although resistivity anisotropy reduces dramatically in the 
pressure released case, it is still present in a narrow temperature region above $T_s$ and below $\sim$130 K, due possibly to
the residual anisotropic strain in the sample [\ref{UPfig2}(c)] \cite{HMan}. Since 
increasing uniaxial pressure enhances both 
the resistivity and spin excitation anisotropy, there must be a direct correlation between the  
resistivity and spin excitation anisotropy.

To further test if spin-spin correlation length also increases at $S_1(1,0)$ but decreases at $S_2(0,1)$ below $T_s$
as expected
from the spin nematic theory \cite{Fernandes2011,rafael,SLiang} 
 [Fig. \ref{UPfig1}(c)]  \cite{Fernandes2011,rafael}, we show in Fig. \ref{UPfig2}(e) temperature dependence 
of the full-width-half-maximum (FWHM) of spin excitations at 10 meV along the marked scan directions at $S_1(1,0)$ and $S_2(0,1)$.  At both wave vectors, we find a clear reduction
of the FWHM in spin excitations around $T_s$. However, since the data collected at $S_2(0,1)$ is along the transverse direction, we cannot directly compare the outcome with spin nematic theory, which predicted an increase in spin-spin excitation correlation length measurable for
scans along the longitudinal direction.

In summary, by using a specially designed in-situ detwinning device to tune the applied uniaxial pressure, we study spin excitation and resistitivity anisotropy in 
AF order and paramagnetic phases of 
BaFe$_{2-x}$Ni$_{x}$As$_{2}$.  For
undoped parent compound BaFe$_2$As$_2$ with $T_N\approx Ts$, we find clear spin excitation anisotropy in the pressure 
released AF ordered phase at 10 meV, but anisotropy vanishes
in the paramagnetic tetragonal phase.  For pressure released 
electron-doped BaFe$_{1.97}$Ni$_{0.03}$As$_2$ with $T_N<T_s$, 
the spin excitation anisotropy at $10$ meV present in the AF ordered phase decreases on warming, persists in the paramagnetic orthorhombic phase ($>T_N$) before vanishing 
in the paramagnetic tetragonal state above $T_s$. 
Assuming the small resistivity anisotropy above $T_s$ in pressure released sample is
extrinsic effect due to residual strain, our results establish a direct correlation between spin excitation and resistivity anisotropy, and 
are consistent with predictions of spin nematic theory \cite{CFang,CXu,Fernandes2011,rafael,SLiang}.  Therefore, our data indicate no additional phase transition above $T_s$, and suggest that the observed resistivity anisotropy in the paramagnetic tetragonal phase \cite{jhchu,matanatar,fisher,chu12,HHKuo,HHKuo15} arises from strong magnetoelastic coupling due to the presence of strong nematic fluctuations.

We are grateful to Sebastien Turc, E. Bourgeat-Lami, E. Leli$\rm \grave{e}$vre-Berna of ILL, France for designing and constructing the detwinning device used at IN8. The neutron scattering work at Rice is supported by the U.S. NSF Grant No. DMR-1700081
(P.D.). The BaFe$_{2-x}$Ni$_{x}$As$_{2}$ single-crystal synthesis work
at Rice is supported by the Robert A. Welch Foundation Grant No. C-1839 (P.D.).

\end{document}